\documentclass{article}
\usepackage{amssymb,graphicx,cancel}
\usepackage[margin=2cm]{geometry}
\usepackage{slashed}
\usepackage{hepunits}
\usepackage{color}
 \usepackage{jheppub}

\newcommand{\be}{\begin{eqnarray}}
\newcommand{\ee}{\end{eqnarray}}

\newcommand{\benum}{\begin{enumerate}}
\newcommand{\eenum}{\end{enumerate}}
\newcommand{\bi}{\begin{itemize}}
\newcommand{\ei}{\end{itemize}}

\begin{document}

\preprint{\\  FERMILAB-PUB-17-520-PPD}

\newcommand\FNAL{Fermi National Accelerator Laboratory, \\ Batavia, IL USA}

\author{Gordan Krnjaic}

\affiliation{\FNAL}

\emailAdd{krnjaicg@fnal.gov} 


\begin{abstract}{ Freeze-in dark matter (DM) mediated by 
 a light ($\ll$ keV) weakly-coupled dark-photon is an important benchmark
 for the emerging low-mass direct detection program.
  Since this is one of the only predictive, detectable freeze-in models, 
  we investigate how robustly such testability extends to other scenarios.
For concreteness, we perform a detailed study of models in which DM couples to 
a light scalar mediator and acquires a freeze-in abundance through Higgs-mediator mixing. 
Unlike dark-photons, whose thermal properties weaken stellar cooling bounds, 
 the scalar coupling to Standard Model (SM) particles is  subject to strong
 astrophysical constraints, which severely limit the fraction of DM that can be produced via freeze-in. 
   While it seems naively possible to compensate for this reduction by increasing the 
   mediator-DM coupling, sufficiently large values eventually thermalize the dark sector with itself and 
  yield efficient DM annihilation to mediators, which depletes the freeze-in population;
only a small window of DM candidate masses near the $\sim$ GeV 
scale can accommodate the total observed abundance. Since many qualitatively similar
issues arise for other light mediators, we find it generically difficult to realize
a viable freeze-in scenario in which production arises only from
renormalizable interactions with SM particles.
We also comment on several model variations that 
may evade these conclusions.
  }
\end{abstract}

\title{Freezing In, Heating Up, and Freezing Out: \\
{  \rm  \Large Predictive Nonthermal Dark Matter and Low-Mass Direct Detection}
}
\date{\today}

\maketitle


\section{Introduction}

Despite overwhelming evidence demonstrating the existence of dark matter (DM) on galactic and cosmological scales \cite{Bertone:2016nfn},
 identifying  its possible non-gravitational interactions remains one of the 
greatest challenges in contemporary physics. 
Historically, much of the DM search program has been guided by the weakly interacting massive particle (WIMP) hypothesis
in which $\sim$ TeV-scale DM carries electroweak charge and acquires its abundance by freezing out of equilibrium from the Standard Model (SM). However, decades of null searches \cite{Roszkowski:2017nbc} have motivated a new, broader paradigm in which DM is the lightest stable SM singlet in a ``hidden sector" with its own dynamics and cosmological history (see \cite{Alexander:2016aln,Battaglieri:2017aum} for a review). In these scenarios the viable DM mass range can span many orders of magnitude, so there is strong motivation for new detection techniques. 
 
Recently, there has been great progress in proposing new direct-detection targets for 
 the $\sim$ keV--GeV DM mass range. Some representative ideas involve scattering off
atomic electrons \cite{Essig:2011nj,Essig:2012yx,Essig:2017kqs}, superconductors \cite{Hochberg:2015pha,Hochberg:2016ajh}, semiconductors \cite{Essig:2011nj,Graham:2012su,Tiffenberg:2017aac,Essig:2015cda}, Helium evaporation \cite{Maris:2017xvi}, Fermi-degenerate materials \cite{Hochberg:2015fth},  Dirac metals \cite{Hochberg:2017wce}, Graphene targets \cite{Hochberg:2016ntt}, magnetic bubble chambers \cite{Bunting:2017net}, color centers \cite{Budnik:2017sbu}, and chemical bonds \cite{Essig:2016crl} (for
 a review, see \cite{Battaglieri:2017aum}). 
Unlike traditional direct detection experiments, designed for WIMP scattering off heavy nuclei 
with $\sim \keV$ recoil energy thresholds, many of these techniques
 exploit observable transitions between small internal energy levels (e.g. atomic ionization) to 
probe much lighter DM,   
which typically deposits $\sim \meV- {\rm few}\, \eV$ per scatter. 
Furthermore, these targets are particularly sensitive to interactions mediated by
 light particles $\phi$ whose recoil spectra 
 are sharply peaked towards low recoil energies, so the scattering cross 
 section can be  greatly enhanced
 \be \label{eq:schematic-sigma}
 \sigma_{ e} \sim \frac{g_{\chi }^2 g_{e}^2 m^2_e }{(q^2 + m_\phi^2)^2} ~ \longrightarrow 
 ~ \frac{g^2_\chi g_e^2 }{ q^4} \sim   5  \times 10^{-38} \, \cm^2 \biggl(    \frac{ g_{\chi}^2 g_{e}^2 }{ 10^{-25}}\biggr) \left(    \frac{ \alpha m_e }{ q  } \right)^4 ,
 \ee
 and may be observable even for extremely small couplings. 
Here $g_{\chi/f}$ are the $\phi$ couplings to dark/visible
 particles where $f$ is a SM fermion and, for illustration, we have taken  $m_\chi \gtrsim m_e$ and normalized to the characteristic
momentum transfer  $q \sim \alpha m_e$  for DM scattering  off atomic electrons \cite{Essig:2011nj}. 
 However, if $g_{f}$ is sufficiently large to thermalize $\phi$ with the SM in the early universe, 
 cosmological bounds from BBN and $\Delta N_{\rm eff}$ generically
 require $m_\phi \gtrsim$ MeV \cite{Pospelov:2010hj}, which is outside the range in which Eq.~(\ref{eq:schematic-sigma}) holds. 
 Thus, in order to exploit this huge enhancement, we  
demand $g_{f} \ll 1$ to avoid thermalizing with the photons, which excludes thermal DM production 
with a sub-MeV mediator, \footnote{A loophole around this argument involves sub-MeV DM with a
comparably light mediator, both of which thermalize with neutrinos after they decouple from the photons \cite{Berlin:2017ftj}. However, such
models must still avoid thermalization with charged particles.} but still allows for non-equilibrium alternatives. 
 
Aside from thermal freeze-out, 
the only other predictive, UV insensitive production mechanism is ``freeze-in" \cite{McDonald:2001vt,Hall:2009bx}, in which DM is not initially populated
during reheating, but produced later through sub-Hubble interactions with SM particles.  Since DM never achieves a thermal number density, its Boltzmann equation is integrable and the final abundance is 
\be\label{eq:freeze-in-naive}
~~ ~~ \Omega_\chi \simeq \frac{m_\chi s_0}{\rho_{\rm cr}}\int_\infty^{m_\chi} \frac{dT}{T} \frac{ n^2_{  \rm \small SM}      }{Hs}  \langle \sigma v_{{ \rm SM}  \to \chi \chi }\rangle  ~~~~~ (\rm traditional~freeze~in),
\ee
where $T$ is the SM temperature, $H$ is the Hubble expansion rate, $s$ is the entropy density, $\rho_{\rm cr} $ is the critical density, $n_{\rm SM}$ is
the number density for some SM species, and a $0$ subscript 
represents a present day value. As with freeze-out,
 the abundance is fixed by the DM-SM interaction and is insensitive\footnote{However, if $\chi$ is produced through 
nonrenormalizable higher-dimension operators whose suppression scale exceeds the highest 
 temperature ever realized in the early universe $T_{\rm max}$, then the abundance in Eq.~(\ref{eq:freeze-in-naive}) will depend 
on UV physics through the ratio $T_{\rm max} / \Lambda$. However, this is never the case for 
the light mediators considered in this paper.} to  the details of earlier, unknown cosmological epochs (e.g. inflation, reheating). 
Although the couplings required for freeze-in production are generically very small for light mediators,
 it may nonetheless be possible to exploit the low thresholds of these experiments to bring the cross section in  Eq.~(\ref{eq:schematic-sigma})
 into the observable range. 

It has been shown that many new direct-detection targets are potentially sensitive 
to the freeze-in parameter space for DM with interactions  mediated by an
 ultra-light ($\ll \keV$) dark-photon \cite{Chu:2011be,Essig:2015cda}. However,
this particular model has many special features, which do not generalize to other mediators (or even to heavier dark-photons).
Most notably, the in-medium thermal suppression of kinetic mixing at finite temperature
 counterintuitively makes the nearly-massless dark-photon {\it safer}
 from astrophysical and cosmological bounds because 
 all on-shell production vanishes in the massless mediator limit \cite{Chang:2016ntp,Hardy:2016kme}. Furthermore, in this limit, the cosmological
  dark-photon production is also suppressed, so it can be neglected in 
 the Boltzmann system and Eq.~(\ref{eq:freeze-in-naive}) can be straightforwardly applied. 
However, in models for which mediator production is not parametrically suppressed in the early 
 universe, Eq.~(\ref{eq:freeze-in-naive}) is no longer valid and
it is necessary to solve the full Boltzmann system that tracks both $\chi$ and $\phi$ densities.
 Similarly, in such models there are nontrivial astrophysical bounds on
the mediator-SM coupling $g_f$ \cite{Chang:2016ntp,Hardy:2016kme}.
Thus, it is important to identify other viable light-mediator 
scenarios that enjoy the enhancement in Eq.~(\ref{eq:schematic-sigma}) and understand 
the generic issues they present. 

Pioneering earlier work \cite{Chu:2011be} studied DM production through renormalizable portals and its subsequent self-thermalization via
dark force interactions. However, this study primarily\footnote{This paper also includes a discussion of heavy scalar DM $S$ with a $Z_2$ 
stabilizing symmetry. In this variation, DM is produced directly through the Higgs portal interaction $S^2 H^\dagger H$ and there is no 
light mediator.} considered a massless dark-photon mediator, with the above-mentioned special properties.
Furthermore, \cite{Knapen:2017xzo} and \cite{Kahlhoefer:2017ddj} recently studied the models and constraints for  light mediators accessible to new direct detection
techniques; however, these analyses leave open the question of whether there exist predictive non-thermal production mechanisms beyond the 
ultra-light dark-photon mediated scenario.  
 
In this paper we extend this discussion by studying a representative benchmark of relevance for low-mass
direct detection: the predictive freeze-in production of dark matter through a light, feebly-coupled scalar mediator $\phi$
with Higgs portal mixing. We find that  even when $g_f \ll 1$ and all SM-DM interaction rates are slower than Hubble, there is generically a  
sizable $\phi$ population, which can thermalize with $\chi$ if $g_\chi$ is sufficiently large. 
 For fixed, viable choices of  $g_f$, there are three qualitatively distinct regimes: 
 \begin{itemize}
 \item {\bf Small Coupling:} If $g_\chi$ is sufficiently small, the hidden sector never thermalizes with itself, so
  hidden-annihilation is always sub-Hubble $\Gamma(\chi \chi \to \phi \phi) \ll H$, and
  production approximates traditional freeze-in. In this regime, the  $\chi$
 abundance increases with $g_\chi$ and Eq.~(\ref{eq:freeze-in-naive}) is approximately valid.   
 \item {\bf Large Coupling:} If  $g_\chi$ is sufficiently large, hidden-annihilation  becomes efficient $\Gamma(\chi \chi \to \phi \phi) \gg H$
and  the hidden sector thermalizes with itself (but still not with the SM). In this regime DM production initially resembles freeze-in, but 
eventually the $\chi$-$\phi$ system reaches thermal equilibrium (at a lower temperature than the SM) and  $\chi$ 
 undergoes an additional phase of {\it freeze-out}.
In this regime, increasing $g_\chi$ merely converts more of the $\chi$ into dark radiation. 
\item {\bf Intermediate  Coupling:} At intermediate values of $g_\chi$ for which the maximum hidden-annihilation rate is {\it briefly} comparable
to Hubble $\Gamma(\chi \chi \to \phi \phi) \sim H$, production interpolates
 between the above regimes and yields a maximum DM abundance for each choice of mass. 
 \end{itemize}
 Intriguingly, we find that for viable mediator-SM couplings $g_f \ll 1$, safe from astrophysical and cosmological bounds, it is generically
  difficult to account for the total freeze-in abundance with only renormalizable couplings to SM particles. Nonetheless, for choices of
  $g_f$ that saturate existing limits, the maximum abundance that can
  be produced is significantly greater than the naive expectation from Eq.~(\ref{eq:freeze-in-naive}) once
  mediator-DM interactions are included in the Boltzmann system.

This paper is organized as follows: section \ref{sec:model} introduces our benchmark model; section \ref{sec:production} describes
the early universe production mechanism; 
section \ref{sec:variations} discusses model variations that evade some of
 our general conclusions; and section \ref{sec:conclusion}
we summarize our findings and suggest future directions of inquiry.


\section{Benchmark Model}
\label{sec:model}

Our  benchmark scenario involves a scalar singlet mediator $\phi$ and mixes with the SM
through the Higgs portal operators 
\be
{\cal L}_{\rm mix} = ( A \phi + \kappa \phi^2) H^\dagger H  , 
\ee
where $A$ and $ \kappa$ are the renormalizable portal couplings. 
After electroweak symmetry breaking (EWSB) the neutral component of the doublet $h$ mixes with $\phi$ and, 
in the small mixing limit, the system is diagonalized with the shift $h \to h+ \sin \theta\, \phi$, so $\phi$ 
acquires mass-proportional couplings to SM fermions. 
 Including a $\phi$ Yukawa interaction with a fermionic DM candidate $\chi$, the 
relevant Lagrangian becomes 
\be\label{eq:phiSMlag}
{\cal L}_{\rm int} = -\phi  \biggl(     g_{\chi}  \overline \chi \chi +  \sum_f  g_f    \overline f f  \biggr) ~,~~ g_f  \equiv \frac{m_f}{v}  \sin\theta  ,
\ee 
where $f$ is a SM fermion of mass $m_{f}$  and $v \simeq  246 \, \GeV$ is the SM Higgs vacuum expectation
value.  
Although this is only one of several possible scalar mediated scenarios, as we will see, 
it captures much of the essential physics and many of the issues encountered here 
apply to a much broader class of variations on this simple setup (see Sec. \ref{sec:variations} for a discussion
and \cite{Kouvaris:2014uoa,Krnjaic:2015mbs} for a complementary studies of thermal freeze-out with the same field content).  

In this model, the cross section for nonrelativistic $\chi e$ scattering is  
\be \label{eq:sigmaechi}
\sigma_{ e}  \simeq   \frac{g_\chi^2  m_e^2  }{ \pi v^2}  \frac{   \sin^2  \!\theta     \,    \mu_{\chi e}^2}{   ( q^2 + m_\phi^2)^2 }
~\longrightarrow~   \frac{g_\chi^2  m_e^2  }{ \pi v^2}  \frac{   \sin^2  \!\theta     \,    \mu_{\chi e}^2}{    (\alpha m_e)^4 }~,
\ee
where $q$ is the three-momentum transfer, $\mu_{\chi e}$ is the reduced mass, and we take the light mediator limit
to highlight the parametric enhancement from Eq.~(\ref{eq:schematic-sigma})
 in terms of $q \sim \alpha m_e$, the characteristic momentum transfer for $\chi$ scattering off 
 atomic electrons \cite{Essig:2011nj}. 
  Our main results (described in Sec. \ref{sec:production} and displayed in Fig. \ref{fig:MoneyPlot}, right panel) 
  are presented in terms of the convention in Eq.~(\ref{eq:sigmaechi}) and are largely independent of the mediator mass so long as 
   $m_\phi \ll \alpha m_e,  m_\chi$.


\begin{figure}[t] \begin{center}
\includegraphics[width=4.8cm]{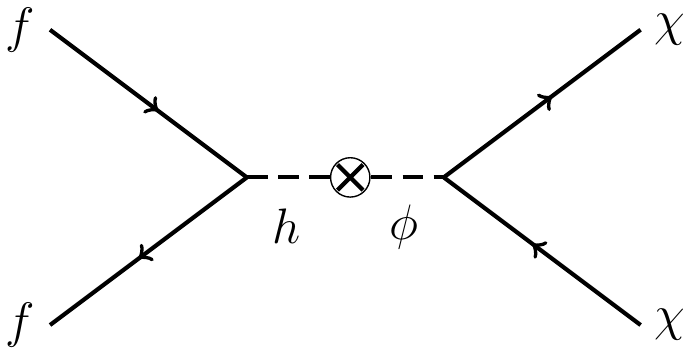} ~~
\includegraphics[width=4.8cm]{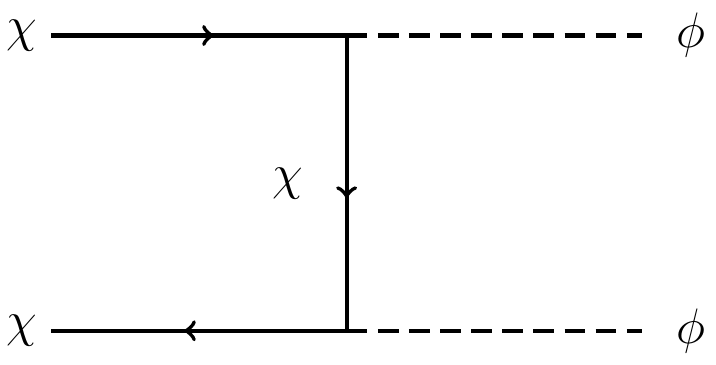}  ~
\includegraphics[width=4.8cm]{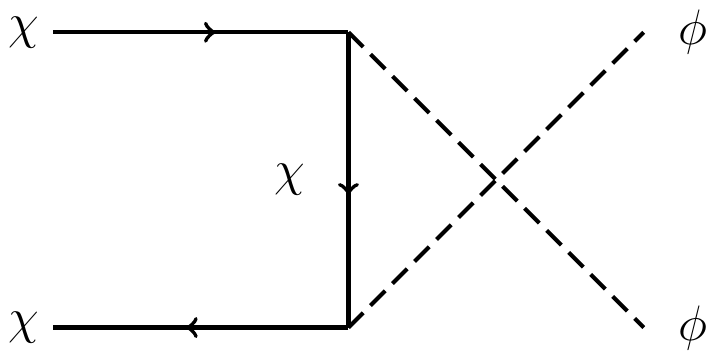}
\end{center}
  \caption{Leading Feynman diagrams giving rise to non-thermal $\chi$ freeze-in production from SM annihilation during the early universe (left) and 
  reannihilation into dark sector mediators $\phi$ (middle, right) once the hidden sector thermalizes with itself. Although the SM initiated process must be slower than hubble to avoid thermalizing
  the dark sector, the   $\chi-\phi$ interactions can achieve chemical equilibrium at a separate temperature set by the energy density 
  transmitted to the dark sector via the left diagram and those depicted in Fig.~\ref{fig:FreezeInDiagram2}.
  }
   \label{fig:FreezeInDiagram}
\vspace{0cm}
\end{figure}


Although our emphasis in this paper is on the freeze-in production for this scenario, it has also been shown that models with 
light scalar mediators $m_\phi \sim 10 \, \MeV$ can yield sufficiently strong DM self interactions $\sigma_{\chi\chi} / m_\chi \sim \cm^2 \gram^{-1}$  
to reduce tension between DM-only N-body simulations and the observed small scale properties of haloes \cite{Kaplinghat:2015aga}. 
It may be interesting to explore whether DM with a scalar mediator in the mass range considered here $m_\phi \ll \alpha m_e$ could preserve
some of these features (as in \cite{Kouvaris:2014uoa} for heavier scalars), but this question is beyond the scope of the present work.


\section{Early Universe Production}
\label{sec:production}


\begin{figure}[t] \begin{center}
\hspace{-0.3 cm}\includegraphics[width=15. cm]{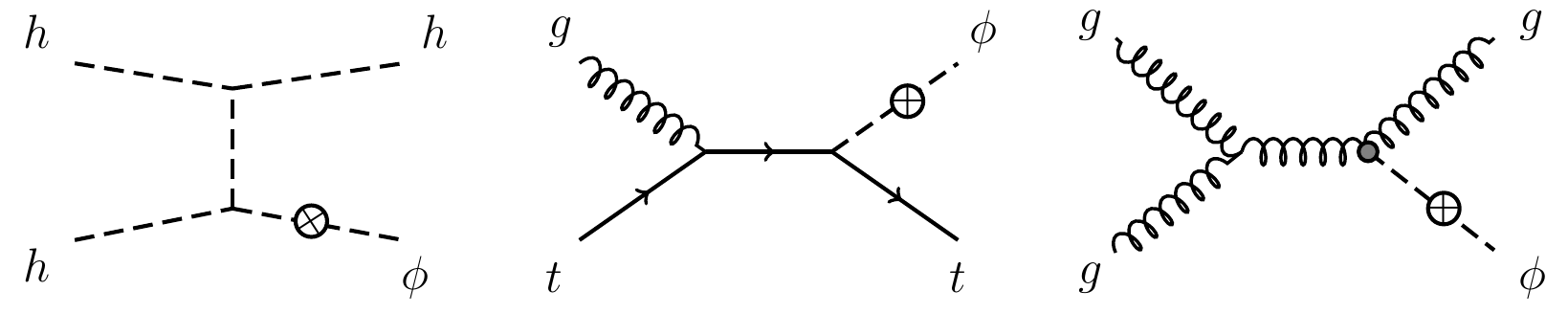} ~~
\end{center}
  \caption{Representative Feynman diagrams giving contributing to $\phi$ freeze in production in the early universe. These processes 
  determine $\rho_\phi$ which (in addition to $\rho_\chi$) determines the hidden sector's temperature $T_D$. The effective vertex 
  in the right diagram  arises from integrating out the top quark for energies in between the weak scale and the QCD confinement scale (see Appendix B).}
   \label{fig:FreezeInDiagram2}
\vspace{0cm}
\end{figure}


\subsection{Boltzmann System}
Our initial condition assumes a hot, radiation dominated universe in the broken electroweak phase.\footnote{It is not strictly necessary to be in the broken phase, but we avoid the complications of Higgs mixing during the electroweak phase transition.} We further assume that  $n_\chi = n_\phi = 0$ after reheating so that sub-Hubble interactions with SM fields are solely responsible for populating the hidden sector.\footnote{For an alternative initial condition in which
a decoupled hidden sector is thermally populated during reheating, but at a different temperature, see \cite{Adshead:2016xxj,Berlin:2016vnh,Berlin:2016gtr}.} The  Boltzmann equation for $\chi$ production  is 
\be \label{eq:full-boltz}
\frac{dn_\chi}{dt} + 3Hn_\chi = \langle \Gamma_{h\to\chi\chi} \rangle n_h  - \langle \sigma v_{\chi \chi \to \rm SM}   \rangle \! \left[  n^2_\chi - n_\chi^{\rm eq}(T)^2\right]
- \langle \sigma v_{\chi \chi \to \phi \phi} \rangle \!      \left[  n^2_\chi - n_\chi^{\rm eq}(T_D)^2\right] \! ,~~~
 \ee
where $H \equiv 1.66 \sqrt{g_*} T^2/m_{\rm \small Pl}$ is the Hubble rate during radiation domination, $m_{\rm Pl} = 1.22 \times 10^{-19} \, \GeV$ is the Planck mass,  and $g_*$ is the number of relativistic degrees of freedom. Here an $^{\rm eq}$ superscript denotes an equilibrium quantity, and we do not assume  equilibrium  between visible and hidden sectors  $(T\ne T_D)$; all relevant SM species are in equilibrium with each other, so we omit their superscript. 
 As discussed above, existing bounds on the mixing angle require that the dark sector never thermalize with the SM in the early universe,
  so $n_\chi \ll n_\chi^{\rm eq}(T)$, but we allow for the possibility that $\chi \chi \longleftrightarrow \phi\phi$ annihilation (shown in the middle and right diagrams of Fig.~\ref{fig:FreezeInDiagram}) equilibrates the hidden 
  sector so that $\chi$ eventually achieves a thermal distribution $n_\chi \to n_\chi^{\rm eq}(T_D)$, where $T_D \ll T$. Thus, we can
  simplify the Boltzmann equation 
\be \label{eq:simple-boltz}
\frac{dn_\chi}{dt} + 3Hn_\chi = \langle \Gamma_{h\to\chi\chi} \rangle n_h  + \langle \sigma v_{\chi \chi \to \rm SM}   \rangle n_\chi^{\rm eq}(T)^2
- \langle \sigma v_{\chi \chi \to \phi \phi} \rangle \!      \left[  n^2_\chi - n_\chi^{\rm eq}(T_D)^2  \right] \! ,~~~
 \ee
where the first two collision terms are independent of $n_\chi$ and serve merely as sources for $\chi$ production. In the limit where
the hidden-sector annihilation rate is also sub-Hubble throughout the production process $\Gamma( \chi \chi \to \phi \phi) \ll  H$, 
Eq.~(\ref{eq:simple-boltz}) recovers the usual freeze-in form in Eq.~(\ref{eq:freeze-in-naive}).


\begin{figure}[t!] 
\hspace{-0.6cm}\includegraphics[width=7.9cm]{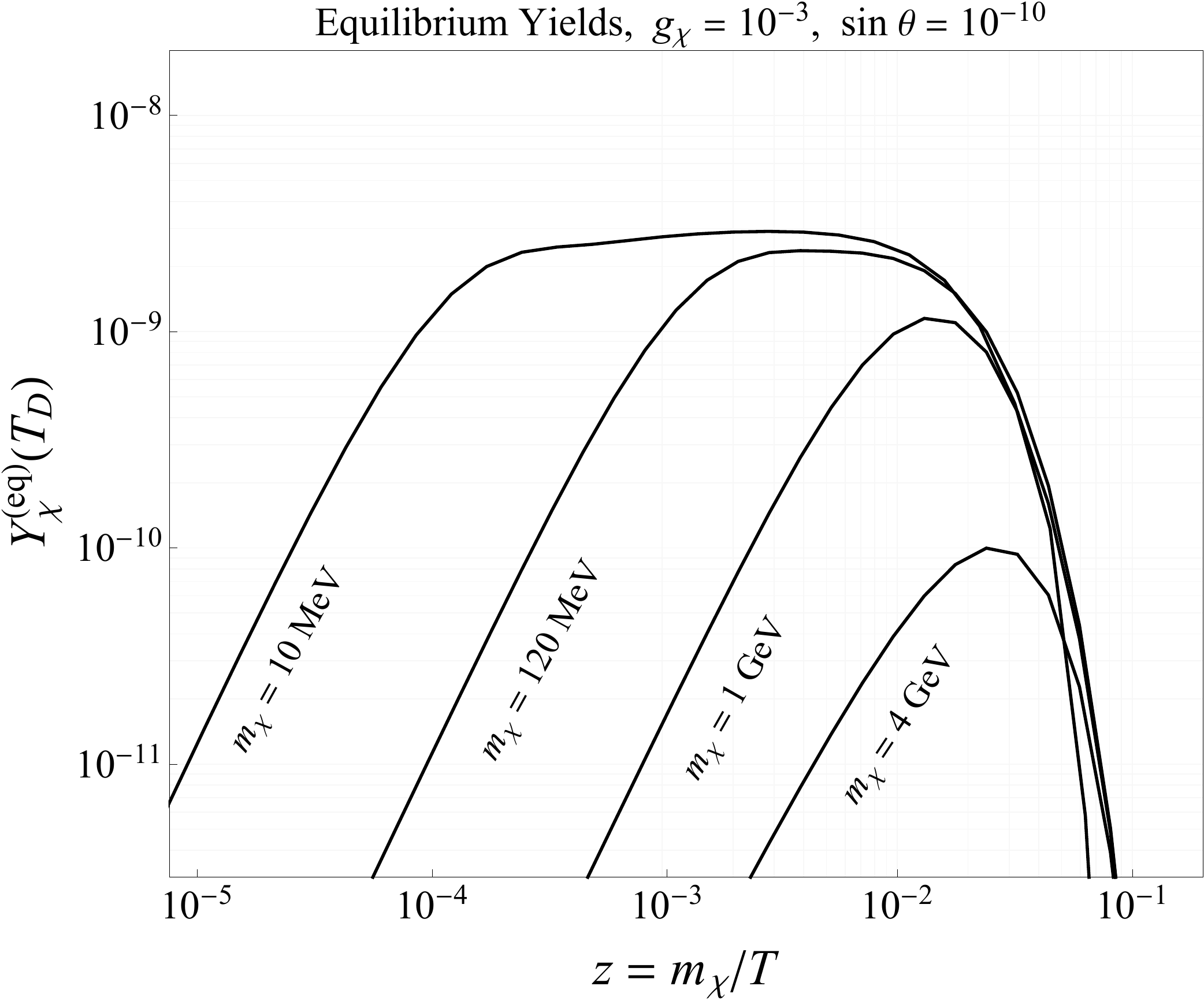}
\hspace{0.3cm}\includegraphics[width=8.6cm]{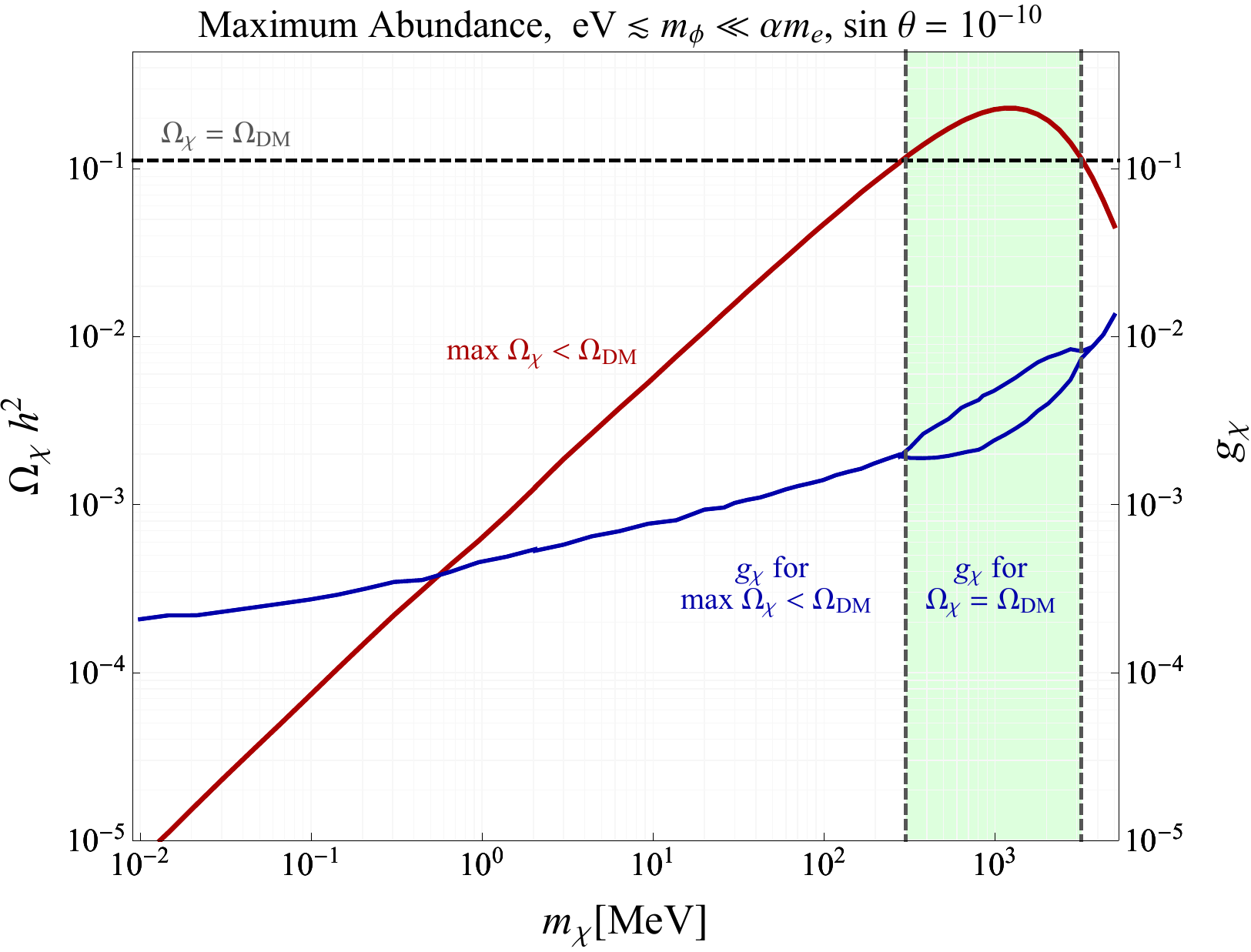}
  \caption{{\bf Left:}  Comoving equilibrium yields $Y^{\rm eq}_\chi = n_\chi^{\rm eq}(T_D)/s(T)$ plotted against
  the dimensionless time variable $z = m_\chi /T$ for a various $m_\chi$. Note that the dark temperature
  is determined primarily by the $\phi$ energy density, which is independent of $m_\chi$, so each mass point 
  has comparable $T_D$ during production; thus, sufficiently large masses are highly Boltzmann suppressed.
  {\bf Right: } Maximum DM abundance for a range of $m_\chi$ with fixed $\sin \theta = 10^{-10}$, which is 
still viable, but near the Red Giant exclusion bound (see left panel of Fig. \ref{fig:MoneyPlot}). Below $m_\chi \lesssim 200$ MeV, 
the blue curve represents the coupling for which the maximum abundance (red curve) can be achieved; larger values merely annihilate
away more of the DM as the number density tracks the equilibrium distributions (left) out to larger values of $z$ (see Fig~\ref{fig:YieldPlot}, left). In
the green band between the dashed vertical lines, ${\rm max} \, \Omega_\chi h^2 > 0.12$ so it is possible to achieve the full abundance for two choices 
of couplings, as shown for a representative mass point in Fig.~\ref{fig:YieldPlot}. 
   }
   \label{fig:EquilibriumPlot}
\vspace{0cm}
\end{figure}


\subsection{Dark Temperature Evolution}

 In this scenario, the hidden sector temperature $T_D$  is defined in terms of the evolving dark sector energy density,  
 which is also produced through SM interactions. The energy transfer Boltzmann equation for direct $\chi$ production is 
\be \label{eq:power-boltz}
 \frac{d\rho_\chi}{dt} + 3 H(P_\chi+\rho_\chi)  =
{\cal P}_{h\to \chi\chi}  n_h  + {\cal P}_{hh \to \chi \chi}     n_h^2 +  {\cal P}_{t\bar t \to \chi \chi}     n_t^2 + {\cal P}_{VV \to \chi \chi}  n_V^2    ,~~~~
 \ee
and the contributions from mediator production (see Fig.~\ref{fig:FreezeInDiagram2}) are 
 \be\label{eq:power-boltz2}
 \frac{d\rho_\phi}{dt}+ 4 H \rho_\phi  =      (  {\cal P}_{t\bar t  \to  g \phi } +   {\cal P}_{t \bar t  \to  h \phi } )   \, n^2_t     
    +          {\cal P}_{tg  \to  t \phi } \, n_t n_g        
+  {\cal P}_{th  \to  q \phi } \, n_h n_t   + {\cal P}_{gg  \to  g \phi } \, n_g^2,   ~~~~
\ee
and we define the thermally averaged quantity
\be 
\label{eq:themal-avg-power}
{\cal P}_{12\to 34}\equiv 
\frac{\int d^3  p_1  d^3  p_2 e^{-\frac{E_1 \!+ \! E_2}{T}}  \Delta E  \sigma_{12\to 34} v_{m \o} }{
\int d^3  p_1  d^3  p_2 e^{-\frac{E_1 \!+ \! E_2}{T}}  } ,~~~~~~~
\ee
where $\Delta E$ is the energy transferred to the hidden sector and 
 $ v_{m \o}$ is
the M\o ller velocity (see Appendix A). Note that the definition of the energy transfer differs depending on whether we 
are considering $\chi$ pair production in Eq.~(\ref{eq:power-boltz}) for which $\Delta E$ tracks both final state particles, or single $\phi$ production in
Eq.~(\ref{eq:power-boltz2}) where it tracks only the energy transferred to $\phi$. 
The combined dark sector energy density $\rho_D = \rho_\phi + \rho_\chi$ defines a dark temperature 
\be \label{dark-temp}
T_D \equiv \left( \frac{30 \, \rho_D}{\pi^2 \xi_{D}   }   \right)^{1/4},
\ee
where $\xi_D$ counts the relativistic degrees of freedom in the hidden sector. Note that 
the dark temperature initially satisfies $T_D = 0$ and
 depends nontrivially on the visible temperature through Eqs.~(\ref{eq:power-boltz}) and (\ref{eq:power-boltz2}), which source its 
 growth in Eq.~(\ref{dark-temp}). 
The dark temperature also determines the equilibrium $\chi$ number density in the hidden sector
\be \label{eq:eq-numberdensity}
n^{\rm eq}_\chi(T_D) = \xi_\chi \int \! \frac{d^3p}{(2\pi)^3} e^{-E/T_D},
\ee
which enters into the RHS of Eq.~(\ref{eq:simple-boltz}). 
In our Boltzmann system we have omitted all processes involving lighter SM fermions, which only contribute negligibly to $\chi$ and $\phi$ production
in the early universe; the dark sector is mainly populated while heavy electroweak states are in equilibrium with the SM thermal bath with
 corrections of order $\sim (m_f/m_t)^2$ for all lighter SM fermions. For this reason, we can safely neglect
longitudinal plasmon mixing effects which can resonantly enhance the $\phi$ production at lower temperatures for which $m_\phi$ is near the plasma frequency  \cite{Hardy:2016kme}; this enhancement does not compensate for the parametric suppression in the coupling relative to continuum production off heavy electroweak states. However, such an effect may be important in model variations in which $\phi$ production arises mainly from its coupling to electrons (see Sec. \ref{sec:variations} for a discussion). Although freeze-in is insensitive to the precise value of the
 reheat temperature, we assume it to be above $m_t$ for all SM electroweak states to be relativistic when dark sector production begins; if the  
 reheat temperature is lower, DM production will proceed through lighter fermions and the resulting abundance will be significantly
  suppressed, but the qualitative features of our discussion are unaffected.
 

\begin{figure}[t!] 
\hspace{-0.5cm}\includegraphics[width=8.0cm]{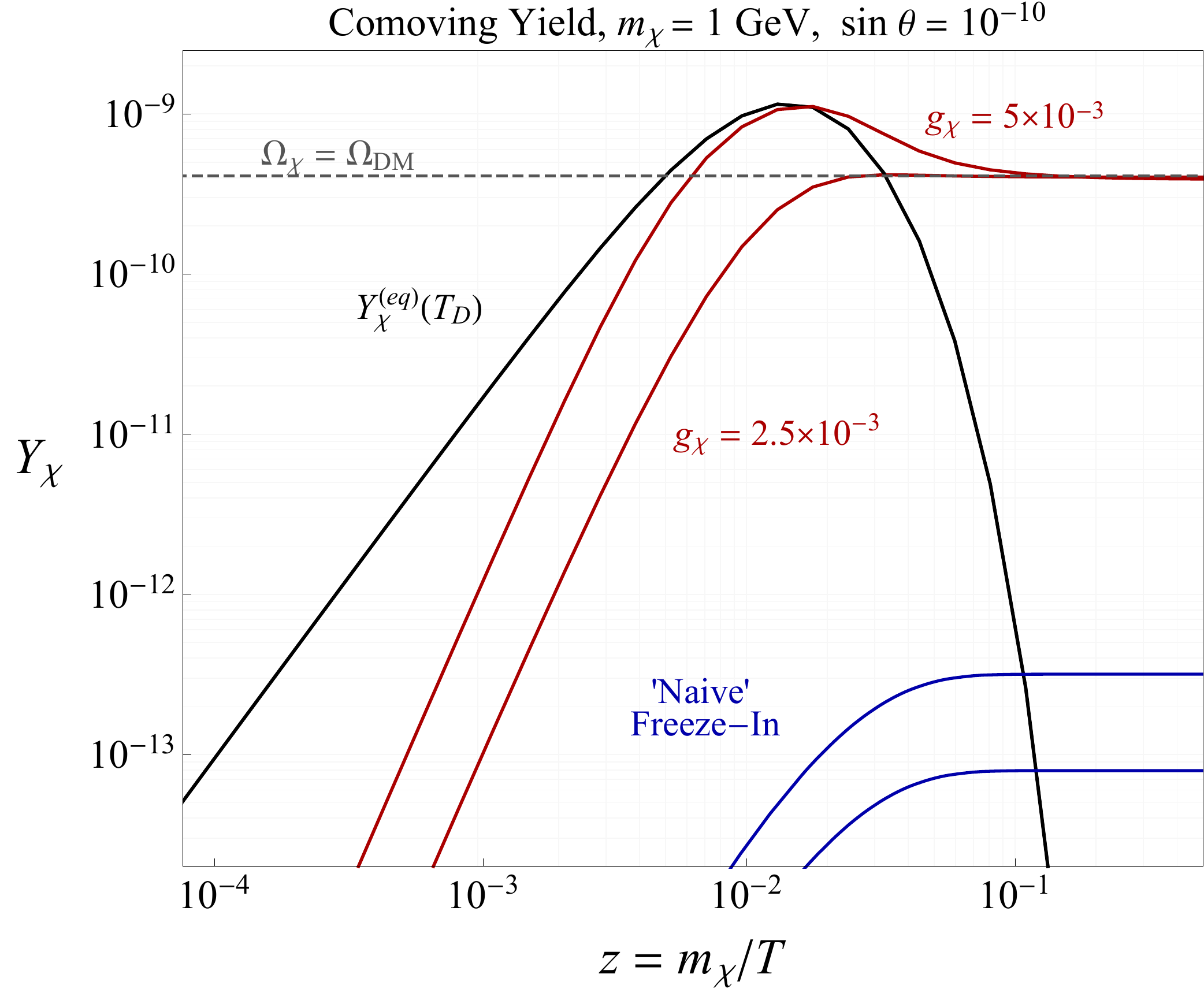}~~~~~~~~
\hspace{-0.5cm}\includegraphics[width=7.75cm]{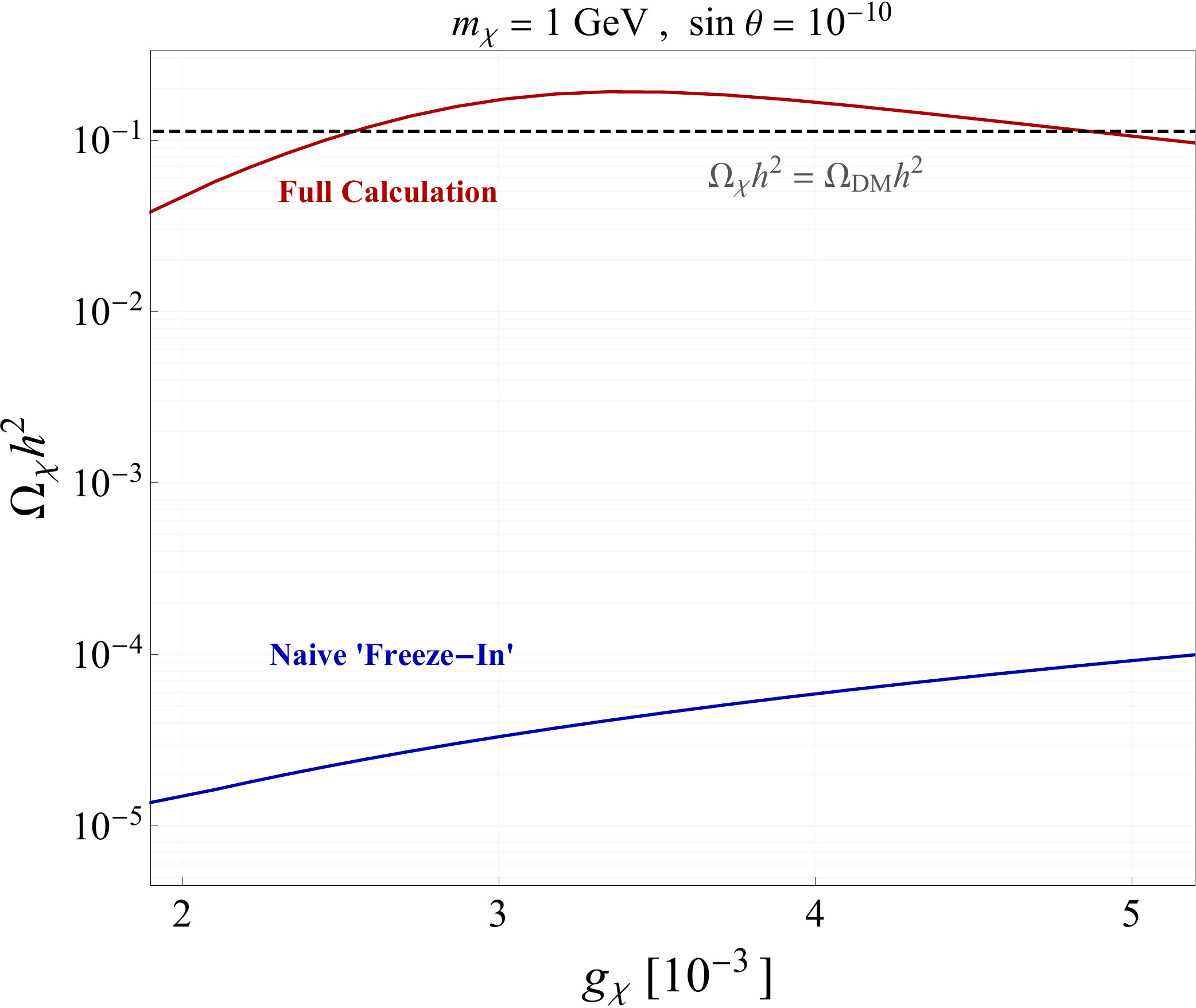}~~~~~~~~
  \caption{{\bf Left:} Comoving yield $Y_\chi = n_\chi(T_D)/ s(T)$ for two representative $g_\chi$, which yield the observed DM
   abundance. Note that for both  $g_\chi < 2.5 \times 10^{-3}$ {\it and }
  $g_\chi > 5 \times 10^{-3}$ the comoving yield is diminished. For reference, the blue curves at the bottom of the plot are the 
  `naive' freeze in result obtained by ignoring the $\chi, \phi$ dynamics in the hidden sector and merely integrating the SM source
  term in the Boltzmann equation from Eq.~ (\ref{eq:freeze-in-naive}). 
   {\bf Right: }  Total $\chi$ abundance for fixed mass and coupling
  plotted alongside different choices of DM self coupling $g_\chi$. Note that the production history for 
  the two points where the red curve intersects the dashed 
  line are represented on the left panel.}
   \label{fig:YieldPlot}
\vspace{0cm}
\end{figure}


\subsection{Numerical Results}


\begin{figure}[t!] 
\hspace{-0.31cm}\includegraphics[width=8cm]{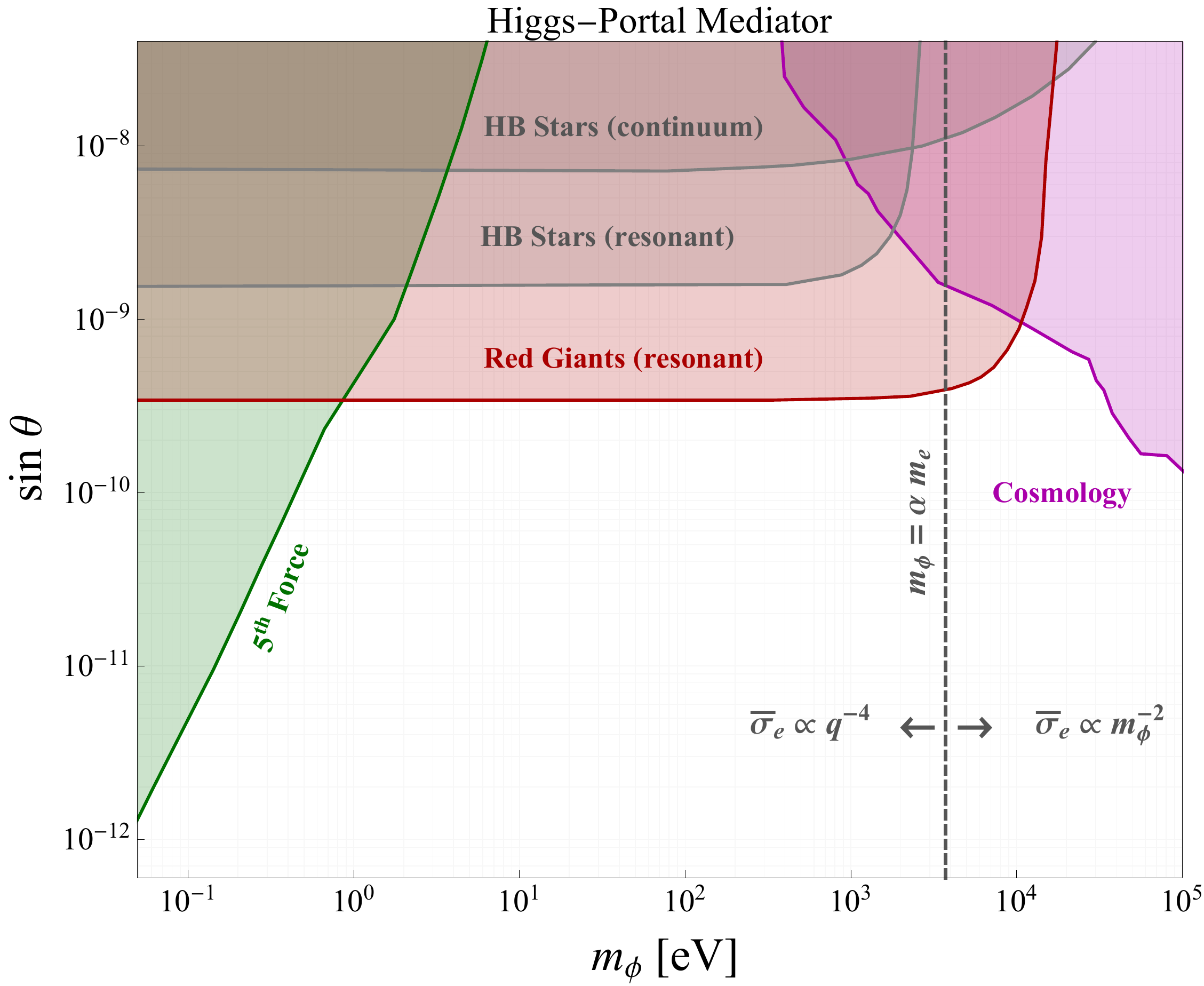}
\hspace{0.21cm} \includegraphics[width=7.8cm]{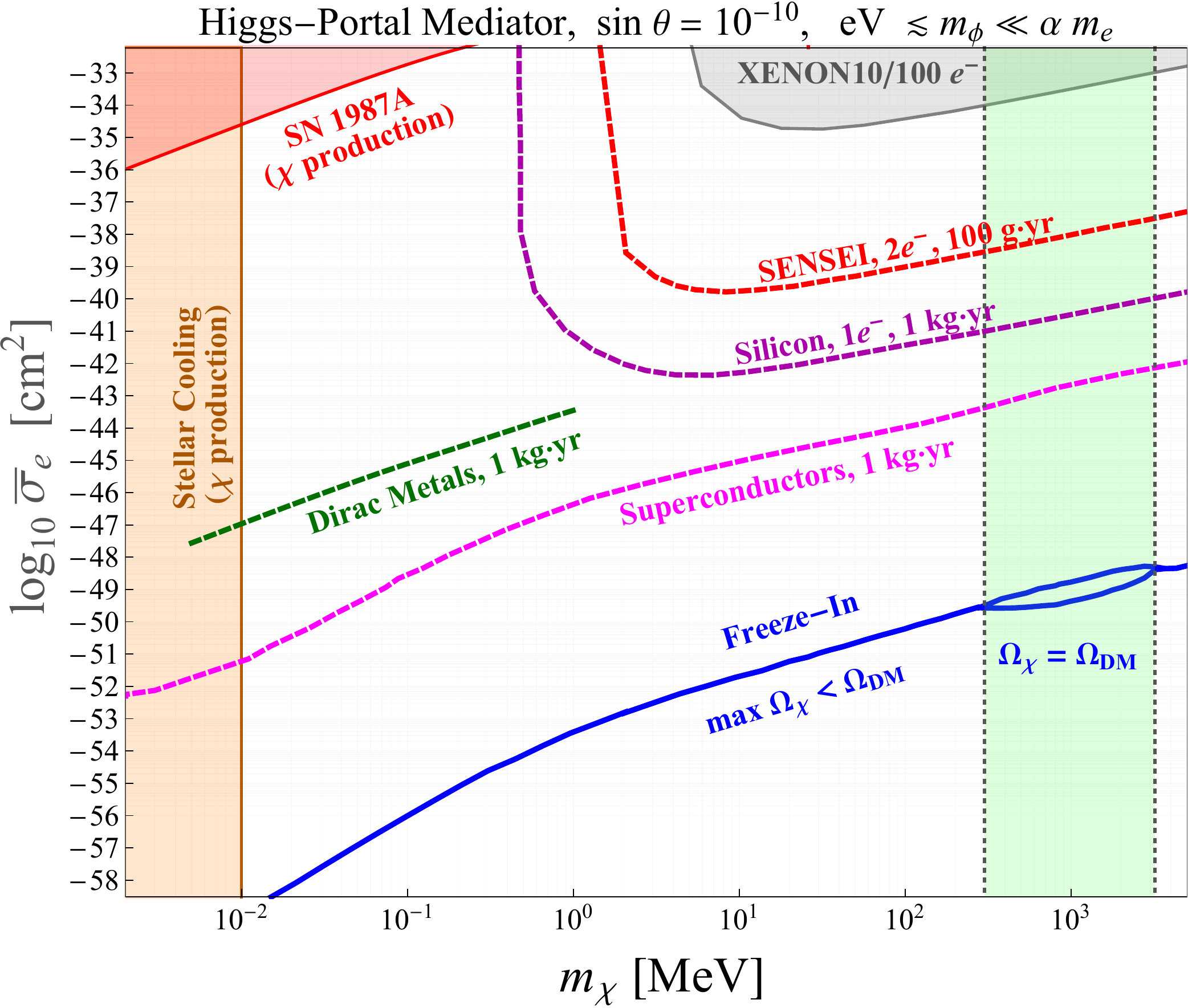}
  \caption{ {\bf Left:} Available parameter space for a light, Higgs-mixed scalar particle plotted alongside
  constraints from 5th forces \cite{Graham:2012su,Kapner:2006si,Geraci:2008hb,Sushkov:2011zz,Decca:2007jq,Hoskins:1985tn}, 
  stellar cooling \cite{Hardy:2016kme}, and cosmology \cite{Flacke:2016szy}. The vertical line represents the approximate boundary 
  where the $\chi e$ scattering rate undergoes a parametric shift for scattering off atomic electrons, which
  is taken to be the reference value for presenting different techniques in \cite{Battaglieri:2017aum}. For other materials with 
  different characteristic momenta, this shift takes place for different values of $m_\phi$.
  {\bf Right:} Constraints and projections for the $\chi e$ cross section 
   for sub-GeV DM. Some representative projections for the SENSEI $2e^-$, Silicon $1e^-$,
and superconductor targets are taken from \cite{Battaglieri:2017aum},
and the Dirac metals projection is from \cite{Hochberg:2017wce}; all projections assume 
$\Omega_\chi \Omega_{\rm DM}$.  
For points outside the green vertical band, the blue curve depicts $ ( {\rm max}\, \Omega_\chi / \Omega_{\rm DM} )\times \overline \sigma_{e} $
  where  sub-Hubble production as described in Sec.~\ref{sec:production} generates the 
   maximum  fractional  abundance (max $\Omega_\chi < \Omega_{\rm DM}$)
    for each $m_\chi$. 
   Inside the vertical green band, the blue curves represent $\overline \sigma_{e}$ 
   evaluated at  $g_\chi$ for which the full
   DM abundance can be accommodated; note there are {\it two} such values
 as depicted in Fig.~\ref{fig:YieldPlot} and on the right panel of Fig.~\ref{fig:EquilibriumPlot}.
   Although the Higgs portal
   mixing $\sin \theta  = 10^{-10}$ may seem ad-hoc, 
   this is chosen as a representative viable value near the stellar cooling
   exclusion region depicted on the left panel of this figure.
  }
   \label{fig:MoneyPlot}
\vspace{0cm}
\end{figure}


For the relevant parameter space considered in this paper $m_\phi \ll \alpha m_e$,
the hidden sector temperature is primarily set by $\rho_\phi$ whose growth rate is proportional to
$\sin^2\theta$ in Eq.~(\ref{eq:power-boltz2}), whereas the collision terms for $\rho_\chi$ in Eq.~(\ref{eq:power-boltz}) all
 depend on $g_\chi^2 \sin^2\theta$ and are further suppressed. Furthermore, $\rho_\phi$ continues
 to accumulate through $gg \to g \phi$ processes long 
after  direct $\chi$ production processes  (e.g. $tt \to \chi \chi$) have frozen out. Thus, the cosmological evolution of  $T_D$ is
insensitive to the DM mass, so sufficiently large values for which
 $m_\chi \ll T_D$ encounter hidden-sector Boltzmann suppression throughout 
their production, $\phi$-thermalization, and re-annihilation phases.  

The left panel of \ref{fig:EquilibriumPlot} presents the equilibrium
$\chi$ yield $Y_\chi^{\rm eq} \equiv n^{\rm eq}_{\chi} (T_D)  /s(T)$ normalized to the total entropy, where $n_\chi^{\rm eq}(T_D)$ is 
computed using Eq.~(\ref{eq:eq-numberdensity}) and depends on the visible temperature implicitly through Eq.~(\ref{dark-temp}).
 Here we see that, as $m_\chi$ increases, the heights of these curves begin to fall, which ultimately limits 
 the total abundance that can be generated for a given mass point; increasing $g_\chi$ increases the abundance in the freeze-in regime, but once 
 the dark sector equilibrates with itself, the yield tracks these equilibrium distributions and eventually diminishes as more of the 
 DM annihilates away into $\phi$ radiation. Thus, for 
 each choice of mass there is a maximum abundance as represented by the red curve in  Fig.~\ref{fig:EquilibriumPlot} (right panel), which
 corresponds to the value of $g_\chi$ from the blue curve. Note that inside the green band, the blue curve splits into two
 segments corresponding to the distinct values that yield the observed DM abundance $\Omega_\chi h^2 \simeq 0.12$, and not the maximum
 abundance (as it does outside this band). 
 
 This counterintuitive behavior can be 
 understood  from the plots in Fig.~\ref{fig:YieldPlot} which shows the production history for a representative mass point $m_\chi = \GeV$
  and $\sin\theta = 10^{-10}$ as the coupling $g_\chi$ is varied. 
 On the left panel, the red curves are solutions to the Boltzmann equation in Eq.~(\ref{eq:simple-boltz}) for two choices of coupling
 that yield the observed DM abundance; the blue curves at the bottom of the plot also show the ``naive" freeze-in result for the same
 choices of $g_\chi$;  as in Eq.~(\ref{eq:freeze-in-naive}), the blue curves are computed by merely integrating the SM collision terms in Eq.~(\ref{eq:simple-boltz}) without including the additional $\chi\chi \longleftrightarrow \phi \phi$ dynamics.  The right panel of Fig.~\ref{fig:YieldPlot} presents the DM abundance as a function of
  $g_\chi$ as computed with the full Boltzmann system (red curve) and  with the naive freeze-in integration (blue curve).

On the right panel of Fig.~\ref{fig:MoneyPlot} we present our main result: the predictive parameter space for 
 $\chi$ production in the early universe plotted in terms of the fiducial $\chi e$ scattering 
 cross section 
 \be
  \overline \sigma_e \equiv \frac{ \mu^2_{\chi e} }{16\pi m_\chi^2 m_e^2} \langle |{\cal  A}(q_0) |^2 \rangle,
 \ee
 where ${\cal  A}(q_0)$ is the matrix element for a free-electron target 
  evaluated at reference momentum $q_0 = \alpha m_e$ following the convention in
  \cite{Battaglieri:2017aum}, which also applies to the other projections shown.
  For the blue curve in this plot, this cross section is given by 
 Eq.~(\ref{eq:sigmaechi}) in the $m_\phi \ll \alpha m_e$ limit and evaluated at $g_\chi$ values, which
either yield the full DM abundance or the largest possible subdominant fraction where this is not possible as shown in Fig.~\ref{fig:EquilibriumPlot} (right); in latter
case the cross section is rescaled by the fractional abundance. The vertical green band in plot 
highlights the mass window for which the total abundance can be achieved and, as in Fig.~\ref{fig:EquilibriumPlot} (right), the blue curve splits into two segments
in the shaded green band to reflect the two values of $g_\chi$ that can  accommodate the total abundance in; the top segment corresponds
to $\chi$ that yield freeze-out in the hidden sector, whereas the bottom segment presents the 
 freeze-in like value. Note that for most of the parameter points shown here max $\Omega_\chi \ll 10^{-2} \, \Omega_{\rm DM}$ (see 
 Fig.~\ref{fig:EquilibriumPlot}, right), so DM self interaction bounds do
 not constrain the viable parameter space; for a discussion of this issue see \cite{Knapen:2017xzo}. Relatedly, there is no 
 perturbative unitary violation in $\chi\chi$ scattering for the $g_\chi$ along this curve since the 
 values that yield the maximum possible abundance for viable choices of $\sin\theta$ are all well below unity (see Fig. \ref{fig:YieldPlot}, right).
 For completeness, we also show the viable parameter space for a Higgs portal mediator on the left panel of Fig.~\ref{fig:MoneyPlot}
 to motivate our benchmarks. 


\section{Model Variations}
\label{sec:variations}
Thus far, we have restricted our attention to the scenario in which 
both $\chi$ and $\phi$ are produced through $h$-$\phi$ mixing during the early universe. We have 
found that the bounds on $\sin \theta \lesssim 5\times10^{-10}$ (see left panel of Fig \ref{fig:MoneyPlot}), 
severely limit the total energy density transferred to the hidden sector, thereby 
making it impossible to achieve the total DM abundance for $m_\chi \lesssim$ few 100 MeV for all values of $g_\chi$. Here 
we consider some possible solutions to this problem that still feature sub-Hubble DM production and
light scalar mediators. 

\subsection{Different SM Current}
\label{subsec:current}
Since much of the discussion above depends on the Higgs-portal coupling relations, it may be
possible to evade some of the above complications with a different pattern of couplings. For the scenario
studied in this paper, the main impediment to generating the observed DM abundance over the $ \eV \lesssim m_\phi \lesssim \alpha m_e$ 
parameter space of interest is the limit $\sin\theta \lesssim5 \times 10^{-10}$, which  is 
driven primarily by resonantly enhanced $\phi$ production through its coupling to electrons in Red Giants (RG) \cite{Hardy:2016kme} -- see
 Fig.~\ref{fig:MoneyPlot} (left). For a minimal Higgs-mixing scenario, the same mixing angle $\theta$ rescales the $\phi$ coupling to electroweak states and thereby 
also suppresses the total energy energy/number density transferred to the hidden sector via electroweak 
production in the early universe. However, the early universe production  
is dominated by the heaviest SM states ($t$, $h$, $W/Z$) at $T\sim v$, so if the $\phi$ coupling to these 
particles can be enhanced relative to the electron coupling in a minimal Higgs-mixing scenario
it is possible to increase the overall abundance at late times. 

\subsubsection{Two Higgs Doublets}
\label{sec:2HDM}
One simple extension, which can accommodate such a variation involves mixing
 $\phi$ with the the neutral CP-even states of a two-Higgs doublet model (2HDM) to break the usual SM relations between quark
and lepton couplings.  For instance, in a Type II 2HDM, $\phi$ could mix with the predominantly up-type doublet, thereby increasing the relative 
ratio of  $t/e$ couplings and enhancing dark sector particle production in electroweak processes.  
However, it is not clear whether a concrete realization of such a scenario can simultaneously 
accommodate the observed DM abundance while satisfying the existing limits on 2HDMs, 
which typically force the $\simeq 125$ GeV CP even scalar close to alignment limit, which
approximately reproduces the usual SM coupling patters \cite{Khachatryan:2014jya,ATLAS:2014kua}, so a careful study is necessary to see if this is possible.

\subsubsection{Vectorlike Fermions}
\label{sec:vectorfermions}
Alternatively, the scalar mediator can couple to heavy vectorlike  
quarks, which are integrated out to induce effective interactions with SM particles  (e.g. $\phi \, G_{\mu\nu}G^{\mu \nu}$, $\phi \, F_{\mu\nu}F^{\mu \nu}$, 
or  $\phi \, \overline qq $ if vectorlike states mix with SM quarks). If the vectorlike quarks are sufficiently decoupled, all dark sector 
production proceeds through effective SM interactions and therefore this class of models makes a firm prediction for the DM-SM 
scattering rate. Furthermore, in this variation there is no tree-level electron coupling, so Red Giant bounds can be significantly weakened \cite{Hardy:2016kme} and more DM  can be produced via freeze-in, but this scenario is harder to test with new detection techniques, most of which
probe DM-electron scattering in various materials.

If, instead, the mediator couples to heavy vectorlike leptons, the IR theory will contain
the $\phi \, F_{\mu\nu}F^{\mu \nu}$ operator (and also $\phi \, \bar \ell \ell$ if the new heavier states mix with right handed leptons). However,
this variation does not ameliorate the Red Giant bounds, which now constrain all early-universe production. Unlike in the Higgs-mixing scenario, where
production off heavier electroweak states in Eqs.~(\ref{eq:power-boltz}) and (~\ref{eq:power-boltz2}) is parametrically enhanced relative to 
the electron scattering cross section, here the production and detection
depend on the same couplings, so 
dark sector production is greatly suppressed. Although the dominant $\phi$ production in this leptophilic scenario
occur at  lower
temperatures $T\sim m_\phi$ where there is a resonant  enhancement from $\phi$--longitudinal-plasmon mixing, 
this effect has been found to be modest (order-few) relative to non-resonant 
continuum production in astrophysical contexts \cite{Hardy:2016kme}, so
it is unlikely to compensate for the large reduction in the overall rate. Nonetheless, a full calculation
of resonant $\phi$ production at these lower temperatures is worth further investigation.\footnote{A related 
calculation in \cite{Fradette:2017sdd} considers heavier $\phi$ production ($m_\phi > 2m_e$) in the Higgs-mixing scenario
for which these resonant enhancements are not important.}

\subsection{Asymmetric Freeze-In}
For the the scenario considered in earlier sections, the abundance is ultimately limited by the additional $\chi \chi \to \phi \phi$
annihilation phase that becomes efficient at depleting the $\chi$ population 
for sufficiently large $g_\chi$. However, if the DM carries a conserved  global
quantum number and has an asymmetric population at late times, then increasing $g_\chi$ will not lead
 to further depletion once all antiparticles are annihilated away. 
 In order to maintain the qualitative features of the scenario considered here (light mediator and sub-Hubble production),
 the freeze-in interaction that produces the DM must also satisfy the Sakharov conditions \cite{Sakharov:1967dj}.
Although models of asymmetric freeze-in have been proposed in the literature \cite{Hall:2010jx},
the asymmetry they produce is suppressed by subtle cancellations \cite{Hook:2011tk} and it is not 
clear whether these limitations can be overcome,
 but this problem deserves further study  (see \cite{Unwin:2014poa} for a discussion).
 
\subsection{Additional Production Source}
One simple way to enhance DM production is to introduce a new source of
$\chi$ or $\phi$ production beyond just the Higgs portal coupling to SM particles in Eq.~(\ref{eq:phiSMlag}).
For instance $\phi$ could also mix with a new, heavy singlet scalar $S$ which thermalizes with the SM in the 
early universe. Since the $S$-$\phi$ mixing angle is a new free parameter, it is possible to increase 
the $\chi$ and $\phi$ abundances via $S$ initiated scattering and decay processes, which still realize freeze-in production if this 
mixing is suitably small. However, this 
modification is no longer predictive and manifestly depends on presently unknown 
UV physics, thereby detracting somewhat from the appeal of the freeze-in mechanism. 

In addition to changing the SM-$\phi$ flavor structure, the SM extensions discussed in Sec. \ref{subsec:current} 
can provide additional production sources if the new heavier particles are populated on shell 
after reheating. New interactions from these states
can enhance $\phi$ and $\chi$ production in the early universe, but
 the magnitude of this enhancement now depends on the
details of the extended sector(s), which adds UV sensitivity to the production mechanism. 



\section{Concluding Remarks}
\label{sec:conclusion}

The recent burst of creativity in devising new techniques sub-GeV direct detection 
may soon make it possible to probe feebly coupled  ``freeze-in" dark matter coupled to a
 very light $\ll \keV$ mediator. 
 However, the only predictive, UV
 complete example in which this has  been demonstrated (DM coupled to an ultra-light dark-photon) has
  subtly special properties that may not generalize to other mediators. 
  To understand how robustly such scenarios can be probed, 
  we have studied the freeze-in production of dark matter
through a light scalar mediator with Higgs portal mixing.

In our analysis we find that the recently updated stellar cooling bounds on light scalars 
have severe, generic implications for any such freeze-in scenario. Most notably, for any sufficiently light ($\ll \keV$) scalar mediator,
compensating for the suppression from this bound requires a large, order-one coupling to dark matter, which quickly thermalizes the dark 
sector with itself. Even though all interactions between dark and visible sectors remain slower than Hubble expansion, the dark sector's 
separate thermal bath enables efficient dark matter annihilation into mediators, thereby depleting its abundance once the mediator--dark-matter coupling
increases beyond the threshold for hidden sector thermalization. Thus, for each choice of dark matter mass and Higgs mixing-angle, there is a maximum cosmological abundance that can be accommodated if the dark sector is only produced through its interactions with the 
Standard Model. We find that for Higgs portal mixing-angles near the limit imposed by astrophysical bounds, it is impossible to generate
the observed dark matter abundance except for a narrow window in the $\sim$ 100 MeV -- few GeV range. Furthermore, even inside this range, 
the cross sections for direct detection off electron targets are several orders of magnitude below future sensitivity projections for proposed experimental
techniques. 
  
We have also considered some possible model variations, which may alter these conclusions and allow for successful freeze-in 
with a light ($\ll$ keV) scalar mediator. In order to produce more DM during the early universe, while satisfying astrophysical bounds (which mostly constrain
the mediator-electron coupling), it may be possible to generalize the minimal Higgs-portal scenario sector to mix the mediator 
with a larger electroweak sector (possibly a 2HDM) and thereby enhance the top-mediator coupling. This modification  
increases early universe freeze-in production while heavy electroweak states are still in equilibrium, but maintains an appreciable coupling
to electrons for direct detection at late times. Similarly, it may be possible
to enhance DM production by preferentially coupling the light mediator to a heavy vectorlike fourth generation of quarks, which are integrated 
out to generate higher dimension operators with SM quarks and gluons.
 Such a may alleviate astrophysical bounds on the mediator coupling and thereby enable viable freeze in production. 
  Other possible variations  include either additional (beyond Standard Model) sources
of mediator or DM production and/or additional CP violation to realize 
a light-mediator variation on asymmetric freeze-in, but these possibilities are beyond the scope of the present work and 
deserve further study.
  

\bigskip

%

\noindent{\it \bf Acknowledgments}: We thank Asher Berlin, Rouven Essig, Roni Harnik, Ciaran Hughes, Seyda Ipek, Yonatan Kahn, Simon Knapen, Robert Lasenby,  Sam McDermott, Gopi Mohlabeng, Maxim Pospelov, and Brian Shuve for helpful conversations. Fermilab is operated by Fermi Research Alliance, LLC, under Contract No. DE- AC02-07CH11359 with the US Department of Energy. 
\medskip


\medskip

\bigskip

\appendix

\section*{Appendix A:   Thermal Averaging}
\renewcommand{\theequation}{A.\arabic{equation}}
\setcounter{equation}{0}

In this appendix we apply the methods used in \cite{Gondolo:1990dk} to compute 
thermal averages for annihilation cross sections and energy transfer rates in the early universe. 

\subsection*{Number Density Collision Terms}
The full Boltzmann equation for DM depletion via  $ \chi(p_1) \chi(p_2) \to A(p_3) B(p_4)$  into SM final states $A$ and $B$ is 
\be
\frac{d n_\chi}{dt} + 3 H n_\chi = C_{n},
\ee
 $H \equiv  \dot a/a = 1.66 \sqrt{g_*} T^2/m_{Pl}$ is the Hubble expansion rate during 
radiation domination and  the collision term for the number density is
\be
       C_{n} =\! \int \! \prod_{i=1}^4  \frac{d^3p_i}{2E_i (2\pi)^3}   (2\pi)^4 \delta^4\bigl( \textstyle{\sum}_j \, p_j \bigr )  
  \langle \,  | {\cal M}|^2  \rangle  \! \bigl[ f^{\rm eq}_A(p_3) f^{\rm eq}_B(p_4)  - f_\chi(p_1) f_\chi(p_2) \bigr] ,
\ee
where $  \langle \,  | {\cal M}|^2  \rangle$ is the spin averaged squared matrix element
and we have omitted bose(pauli) enhancement(blocking) factors. Using detailed balance $f^{\rm eq}_\chi (p_1) f^{\rm eq}_\chi (p_2) = 
f_A^{\rm eq}(p_3)f^{\rm eq}_B(p_4)$ and following the derivation in   \cite{Gondolo:1990dk}, the Botlzmann equation becomes 
\be
\frac{d n_\chi}{dt} + 3 H n_\chi = -\langle \sigma v_{\chi \chi \to A B}\rangle \left[   n_\chi^2  - (n^{\rm eq}_\chi)^2      \right],
\ee
where the thermally averaged annihilation cross section is defined to be 
\be
\langle  \sigma v_{\chi \chi \to { AB}} \rangle  \equiv \frac{1}{8 m_\chi^4 T K_2\bigl( \frac{m_\chi}{T}\bigr)^2} \int_{4m_\chi^2}^\infty ds \,  \sigma_{\chi \chi \to \phi \phi} \sqrt{s} (s - 4m_\chi^2) K_1\biggl( \frac{\sqrt{s}}{T} \biggr),~~~~
\ee
where $K_{1,2}$ are Bessel functions of the first and second kinds, respectively. 

In the traditional ``Freeze In" regime, the initial condition is $n_\chi(0) = 0$ and the $\chi$ production rate in SM annihilation processes is slower than Hubble expansion, so
the $n^2_\chi$ term can be neglected and the Boltzmann equation becomes integrable
\be
\frac{d n_\chi}{dt} + 3 H n_\chi = \langle \sigma v_{\chi \chi \to A B}\rangle  (n^{\rm eq}_\chi)^2      ,
\ee
and we can define comoving yields $Y_i \equiv n_i/s$  and change the time variable to $z \equiv m_\chi /T$ to obtain
\be
\frac{d Y_\chi}{dz}  =  \frac{  \langle \sigma v_{\chi \chi \to A B}\rangle s }{Hz } (Y^{\rm eq}_\chi)^2 ,
\ee
which can be integrated to obtain the asymptotic abundance at late times 
\be
 \Omega_\chi = \frac{m_\chi s_0 }{\rho_{\rm cr}} \,  Y_\chi(\infty)  =  \frac{m_\chi s_0}{\rho_{\rm cr}}\int_0^\infty \frac{dz}{H z } s \langle \sigma v_{\chi \chi \to A B}\rangle   (Y^{\rm eq}_\chi)^2,
\ee
where $s_0 \simeq 2969 \cm^{-3}$ is the present day CMB temperature, $\rho_{\rm cr} = 8.1 h^2 \times 10^{-47} \,\GeV^{4}$ is the critical density. 

\subsection*{Energy Density Collision Terms}
Here we generalize the argument in \cite{Gondolo:1990dk} to calculate the thermally averaged energy transferred to particle $1$ (which we will later identify with $\chi$ or $\phi$)
 in  a  $1+2 \longleftrightarrow 3 + 4$ process governed by the Boltzmann equation
 \be
 \frac{d\rho_1}{dt} + 4 H(P_1 + \rho_1) = C_\rho,
 \ee
   Ignoring quantum statistical factors and assuming the other particles are in equilibrium  with the radiation bath, the 
collision term is 
\be
       C_{\rho} =\! \int \! \prod_{i=1}^4  \frac{d^3p_i}{2E_i (2\pi)^3}   (2\pi)^4 \delta^4\bigl( \textstyle{\sum}_j \, p_j \bigr )  
  \langle \,  | {\cal M}|^2  \rangle   E_1  \bigl[ f_3^{\rm eq}(p_3) f^{\rm eq}_4(p_4)  - f_1(p_1) f_2^{\rm eq}(p_2) \bigr] ,
\ee
where we make no assumption about particle $1$ being in thermal equilibrium with the others. 
Using detailed balance $f_3^{\rm eq}(p_3) f_4^{\rm eq}(p_4) = f_1^{\rm eq}(p_1) f_2^{\rm eq}(p_2)$ and
the definition of the cross section  
\be
\sigma_{12  \to 34} = \frac{1}{ 4 F } \int \frac{d^3 p_3}{2E_3 (2\pi)^3}
\frac{d^3 p_4}{2E_4 (2\pi)^3}      (2\pi)^4 \delta^4\bigl( \textstyle{\sum}_j \, p_j \bigr )  
  \langle \,  | {\cal M}|^2  \rangle,  \!      
\ee
where $F = \sqrt{(p_1\cdot p_2)^2 - m_1^2 m_2^2} $ is the usual flux factor, so we get 
\be
C_\rho = \! \int \!    \frac{d^3p_1}{2E_1 (2\pi)^3}    \frac{d^3p_2}{2E_2 (2\pi)^3}        4 F \sigma_{12 \to 34} 
   E_1  f^{\rm eq}_2(p_2)     \bigl[ f^{\rm eq}_1 (p_1)   - f_1(p_1) \bigr] .
\ee
Since the  production/absorption rate is always slower than Hubble expansion for our purposes, 
the $f_1$ term can be neglected, so we get 
\be
C_\rho = \! \int \!    \frac{d^3p_1}{2E_1 (2\pi)^3}    \frac{d^3p_2}{2E_2 (2\pi)^3}        4 F \sigma_{12 \to 34} \,
   E_1     f^{\rm eq}_1 (p_1)   f^{\rm eq}_2(p_2)  =  {\cal P}_{12\to 34} \, n^{\rm eq}_2  n^{\rm eq}_1 ,
\ee
where the thermally averaged power transfer rate is
\be 
\label{eq:themal-avg-power}
{\cal P}_{12\to 34}\equiv 
\frac{\int d^3  p_1  d^3  p_2 e^{-\frac{E_1 \!+ \! E_2}{T}}  E_1 \sigma_{12\to 34} v_{m \o} }{
\int d^3  p_1  d^3  p_2 e^{-\frac{E_1 \!+ \! E_2}{T}}  } ,~~~~~~~
\ee
and where the M{\o}ller velocity is $v_{m \o} \equiv F /E_1 E_2$. the thermally averaged power is  and denominator factor can be evaluated analytically 
\be \label{eq:thermal-avg-denominator-mixed}
\!\! \int  \! d^3  p_1  d^3  p_2 e^{-\frac{E_1 \!+ \! E_2}{T}} \!= (4 \pi m_1 m_2 T)^2 K_2\! \left( \frac{  m_1}{T}\!\right) K_2 \!\left( \frac{  m_2}{T}\!\right).~~~~~~~
\ee
To evaluate the numerator of  Eq.~(\ref{eq:themal-avg-power}), we follow the procedure in \cite{Gondolo:1990dk}  we perform all trivial angular integrations and perform the coordinate transformation $(E_1, E_2 \cos \theta) \to (E_+, E_-, s)$ where $\theta$ is the CM angle between the incoming three-vectors, 
$E_{\pm} = E_+ \pm E_-$, and $s = (p_1 + p_2)^2$ to get
\be
\int d^3  p_1  d^3  p_2 e^{-\frac{E_1 \!+ \! E_2}{T}}  E_1 \sigma v_{m \o} 
  &=&2  \pi^2 \! \int_{s_0}^\infty ds \int^\infty_{\sqrt{s}}  dE_+ \int_{-\varepsilon}^{\varepsilon} dE_-     \,  e^{-E_+/T}   \sigma _{12\to 34} E^2_1 E_2  v_{m \o}   \nonumber \\ 
 &=& 2 \pi^2 T  \! \int_{s_0}^\infty  \! ds \, \sigma_{12\to 34} s \sqrt{1- \frac{s_0}{s}} K_2\! \left(\frac{\sqrt{s}     }{T}\right)       ,
\ee
where we have used  $\varepsilon = \sqrt{1 - s_0/s} \sqrt{E_+^2-s}$, so the power transfer becomes 
\be \label{eq:final-power-tansfer}
{\cal P}_{12\to 34}
= \frac{1}{      8 m^2_1 m^2_2  T K_2\! \left( \frac{  m_1}{T}\!\right) K_2 \!\left( \frac{  m_2}{T}\!\right)    } \! \int_{s_0}^\infty  \! \! ds \, \sigma_{12\to 34}  \,  s F \sqrt{1- \frac{s_0}{s}} K_2\! \left( \! \frac{\sqrt{s}}{T}\right),
\ee
which is valid even in the $m_i \to 0$ limit provided that the Bessel functions are properly expanded to cancel the mass
dependence in the prefactor. Using detailed balance, we can relate the resulting collision term to the reverse process 
\be
{\cal P}_{12 \to 34} \, n_1^{\rm eq} n_2^{\rm eq} = {\cal P}_{34 \to 12} \,   n_3^{\rm eq} n_4^{\rm eq},
\ee
which describes the SM+SM $\to$ SM $\phi$ processes in Eq.~(\ref{eq:power-boltz2}) where we identify $\phi$ with particle 1 of this argument.

For processes in which 2$\chi$ particles are produced via SM annihilation, a similar argument yields the expression 
\be \label{eq:power-tansfer-chi}
{\cal P}_{\chi \chi \to AB} = 
  \frac{1}{8 m_\chi^4 T K_2\bigl( \frac{m_\chi}{T}\bigr)^2} \int_{4m_\chi^2}^\infty ds \,  \sigma_{\chi \chi \to AB}  \, s (s - 4m_\chi^2) K_2\biggl( \frac{\sqrt{s}}{T} \biggr),~~~~
\ee
which tracks the energy transfer to both hidden sector particles produced in this process. whose form is appropriate for the collision terms in  Eq.~(\ref{eq:power-boltz}).


\appendix

\section*{Appendix B:   Cross Sections}
\renewcommand{\theequation}{B.\arabic{equation}}
\setcounter{equation}{0}

\subsection*{DM Hidden Sector Annihilation $\chi \chi \to \phi \phi$ }

In the hidden sector, the $\chi \chi \to \phi \phi$ annihilation cross section is 
\be
\sigma_{\chi \chi \to \phi \phi} =  \frac{g_\chi^4}{16 \pi s^2 (s - 4 m_\chi^2)     }  \left[   (s^2 - 16 m_\chi^2 s - 32 m_\chi^4 ) \coth^{-1} \! \beta_\chi^{-1}   - (s- 8m_\chi^2) \sqrt{s(s - 4m_\chi^2)} \, \right] ,~~~~~~       
\ee
where we have taken the limit $m_\chi \gg m_\phi$, which is appropriate for the full range of parameters 
we consider in this work. The thermal average has the familiar form, but evaluated at the hidden sector 
temeprature
\be
\langle  \sigma v_{\chi \chi \to \phi \phi} \rangle  = \frac{1}{8 m_\chi^4 T K_2\bigl( \frac{m_\chi}{T_D}\bigr)^2} \int_{4m_\chi^2}^\infty ds \,  \sigma_{\chi \chi \to \phi \phi} \sqrt{s} (s - 4m_\chi^2) K_1\biggl( \frac{\sqrt{s}}{T_D} \biggr),~~~~
\ee
which depends on the visible temperature through Eqs.~(\ref{eq:power-boltz}) and (\ref{dark-temp}).

\subsection*{Direct DM Production $ h \to \chi \chi $  }

\be
\Gamma_{h \to \chi \chi} = \frac{g_\chi^2 \sin^2\!\theta \,  m_h}{ 8 \pi } \left( 1 - \frac{4m_\chi^2}{m_h^2} \right)^{3/2},~~
\ee

\subsection*{Direct DM Production $ \chi \chi \to ff$  }
The cross section for $\chi$ annihilation into SM fermions $f$ is given by
\be
\sigma_{\chi \chi \to ff} = \frac{g_\chi^2 \sin^2\! \theta   \,  m_f^2    }{16 \pi v^2 s}  
   \beta_f^3 \, \beta_\chi ~,~~
\ee
where we have defined $\beta_i \equiv \sqrt{1 - 4m_i^2/s}$

\subsection*{Direct DM Production: Higgs Annihilation $ \chi \chi \to hh$  }
\be
\sigma_{\chi \chi \to hh} = \frac{9 g_\chi^2 \sin^2\! \theta   \,  m_h^4     }{32 \pi v^2 s^2}  
\left(1-\frac{   4m_\chi^2}{s} \right)  
     \sqrt{ \frac{ s - 4m_h^2 }{  s - 4m_\chi^2     }  } ~,~~
\ee

\subsection*{Direct DM Production: Vector Annihilation $ \chi \chi \to VV$  }

\be
\sigma_{\chi \chi \to VV } = \frac{ \, g_\chi^2 \sin^2\! \theta   \,  m_h^4     (s- 4m_\chi^2)(12 m_V^4 - 4 m_V^2 s  + s^2)     }{72 \pi v^2 s^3}  
     \sqrt{ \frac{ s - 4m_V^2 }{  s - 4m_\chi^2     }  } ~,~~
\ee

\subsection*{Mediatior Production: Electroweak Scalar Annihilation $ ff \to h \phi$  }    
For each fermion $f$ there are three contributions to the 
$f(p_1)\bar f(p_2) \to \phi(p_3) h(p_4)$ process 
\be
{\cal M}  (f\bar f \to \phi h) = - \sin\theta \,  \overline u(p_1) \left(   \frac{m_f^2}{v^2}       \frac{  (\displaystyle{\not}{p_1}  - \displaystyle{\not}{p_3}  + m_f)  }{t-m_f^2} 
 +
 \frac{m_f^2}{v^2}  \frac{(\displaystyle{\not}{p_1} - \displaystyle{\not}{p_4}  +  m_f)  }{u-m_f^2}  
  +
   \frac{3 m_h^2 m_f }{v^2}  \frac{1}{s - m_h^2}    \right) v(p_2),~~~~~~~
\ee
where we have also included the $s$-channel diagram involving the trilinear scalar interaction. 
We calculate the total cross section using FeynCalc \cite{Shtabovenko:2016olh} 

\subsection*{Mediatior Production: Electroweak Scalar Compton $ fh \to f \phi $  } 
As above, the higgs-fermion scattering process $h(p_1)f(p_2) \to \phi(p_3)f(p_4)$  is represented by three Feynman diagrams with amplitude
\be
{\cal M}  (hf \to \phi f) =  - \sin\theta \,  \overline u(p_2) \left(   \frac{m_f^2}{v^2}       \frac{  (\displaystyle{\not}{p_1}  + \displaystyle{\not}{p_2}  + m_f)  }{s-m_f^2} 
 +
 \frac{m_f^2}{v^2}  \frac{(\displaystyle{\not}{p_2} - \displaystyle{\not}{p_3}  +  m_f)  }{u-m_f^2}  
  +
   \frac{3 m_h^2 m_f}{v^2}  \frac{1}{t - m_h^2}       \right) u(p_4).~~~~~~~
\ee
We calculate the total cross section using FeynCalc \cite{Shtabovenko:2016olh} where we sum over colors
and average over spins.

\subsection*{Mediatior Production: Higgs Semi Annihilation $ hh \to h \phi$  }
The cross section for $\phi$ production via Higgs semi annihilation  is 
\be
\sigma_{hh\to h \phi} = \frac{ 9 m_h^4 \sin^2\!\theta   }{ 16 \pi v^4  (s-4m_h^2)(s -m_h^2)  }  
\left[  (s + 2m_h^2 )(s + 8m_h^2 ) \beta_h   + 24 m_h^2 (s - m_h^2) \tanh^{-1}    \beta_h \,   \right],~~~~~
\ee
where $\beta_h \equiv \sqrt{1 - 4m_h^2/s}$ and we have taken the $m_\phi \to 0$ limit, appropriate for the early universe while the Higgs is still
part of the thermal bath.

\subsection*{Mediatior Production: QCD Annihilation $ ff \to g \phi$  } 
Our main focus will be annihilation off SM top quarks with cross section
\be
\sigma_{f\bar f \to g \phi} = \frac{4 \alpha_s  \sin^2\!\theta \,m_f^2 }{ s  (s - 4m_f^2)   v^2  }  \left[  (s+4 m_f^2 )\tanh^{-1}  \beta_f      - (s - 2m_f^2) \beta_f     \right]   ,~~~~
\ee
 where we have summed (not averaged) over all colors to keep track of 
 all degrees of freedom that enter into the collision term of the Boltzmann equations. 

 \subsection*{Mediatior Production: QCD Compton $ fg \to f \phi$  }
For the QCD compton process, we have 
\be
\sigma_{fg \to f \phi} = \frac{ \alpha_s  \sin^2\!\theta   \,m_f^2   }{    2s^2   (s - m_f^2)^3   v^2    }     \left[    
  (s-m_f^2 ) (3s + m_f^2) (m_f^4 - 8 m_f^2 s  - s^2)  +
  2s^2(s+3m_f^2)^2 \log \frac{s}{m_f^2} 
 \right] ,~~~~~
\ee
where we have summed (not averaged) over colors to account for all relevant degrees of freedom in
the thermal bath.

\subsection*{Mediatior Production: Glue Scatter $ gg \to g \phi$  }
For $ m_t \gg T \gg m_\phi$, we can integrate out the top quark and evaluate the scattering 
process $gg\to g\phi$ which remains effective until confinement
around $T \sim \Lambda_{\rm QCD} \approx 200\, \MeV$. Neglecting $m_\phi$, the differential rate for this 
process is 
\be 
\frac{d\sigma_{gg\to g\phi}}{dt} =  -\frac{\pi^2 \alpha_s^3 (s^2 + st + t^2 )^2 }{12 v^2 s^3 t  (s+t) },
\ee
where again we have summed (but not averaged) over all colors. Note that the RHS is always positive 
in the physical phase space where $t < 0$ and that there is a $t$-channel, forward scattering
 singularity as $t \to 0$, which is regulated by the QCD Debye mass $m^2_g(T) = 8 \pi \alpha_s T^2$ at 
 finite temperature.  

\bibliographystyle{JHEP}

\bibliography{scalarfreezein}

\end{document}